\documentclass[12pt]{article}

\newcommand{\m}{\mathrm}
\newcommand{\be}{\begin{equation}}
\newcommand{\ee}{\end{equation}}
\newcommand{\ba}{\begin{eqnarray}}
\newcommand{\ea}{\end{eqnarray}}

\usepackage{graphicx}
\usepackage{amssymb}
\usepackage{amsmath}
\usepackage[T1]{fontenc} %for \boldsymbol
\usepackage[ansinew]{inputenc} %for \boldsymbol
\usepackage[nosort]{cite}
\newcommand{\inbar}{\vrule height1.57ex width.4pt depth0pt}
\newcommand{\SW}{\relax{\hbox{$\ \inbar\kern-.285em{\rm S}$}}}

\begin{document}
\thispagestyle{empty}
\begin{center}

\null \vskip-1truecm \vskip2truecm

{\Large{\bf \textsf{Characterising the Most Rapidly Rotating AdS$_5$-Kerr Black Holes}}}

{\large{\bf \textsf{}}}

{\large{\bf \textsf{}}}

\vskip1truecm

{\large \textsf{Brett McInnes}}

\vskip1truecm

\textsf{\\  National
  University of Singapore}

\textsf{email: matmcinn@nus.edu.sg}\\

\end{center}
\vskip1truecm \centerline{\textsf{ABSTRACT}} \baselineskip=15pt
\medskip
Classical Censorship permits AdS$_5$-Kerr black holes with arbitrarily large angular momenta per unit mass, which does not seem reasonable from a holographic point of view. However, it has been shown that, when these black holes are embedded in string theory, their angular momentum per unit mass is in fact bounded by $2\sqrt{2}L$, where $L$ is the asymptotic curvature scale. One might suppose that the most rapidly rotating AdS$_5$-Kerr black holes, with angular momentum per unit mass close to this bound, might be unstable, for example, to a superradiant instability. We show that this is not always true: there is a small domain in the AdS$_5$-Kerr parameter space corresponding to black holes which are stable against superradiance and yet nearly saturate the stringy bound.

\newpage

\addtocounter{section}{1}
\section* {\large{\textsf{1. Can AdS$_5$-Kerr Black Holes Rotate Arbitrarily Quickly?}}}
It is well known that, in spacetime dimensions higher than five, Cosmic Censorship does not forbid arbitrarily large black hole angular momenta for a given mass \cite{kn:reall}. In the five-dimensional, asymptotically flat case, Censorship does have that effect, as of course it does in four dimensions \cite{kn:kerr}.

In the study of holographic duality \cite{kn:casa,kn:nat,kn:bag}, asymptotically AdS five-dimensional black holes play a central role, so it is natural to extend the question to ask: in the AdS$_5$-Kerr case, does (classical) Censorship put a bound on the angular momentum of a black hole\footnote{For the sake of simplicity and clarity, we shall throughout this work consider only singly-rotating AdS$_5$-Kerr black holes. For a discussion of Censorship for singly rotating higher-dimensional black holes in the asymptotically flat case, see \cite{kn:dad}.} of given mass?

The answer to this question is that classical Censorship \emph{does not} impose such a bound \cite{kn:96}; one can construct explicit examples of such black holes, of given mass, which satisfy classical Censorship while having an arbitrarily large angular momentum to mass ratio $\mathcal{A}$. Indeed, the effect of classical Censorship for these black holes is just to exclude a \emph{band} of values for $\mathcal{A}/L$ (where $L$ is the asymptotic curvature length scale) around unity. That is, $\mathcal{A}/L$ has \emph{either} to be smaller than a certain value less than unity, \emph{or} greater than a certain value larger than unity.

The gauge-gravity duality, which motivates our interest in AdS$_5$-Kerr spacetimes, implies that these objects are dual to non-gravitational strongly coupled systems with non-zero angular momentum densities. However, one would expect \emph{any} fluid with a given energy density to become unstable if its vorticity can be increased without bound. Thus it seems that Cosmic Censorship, which as we have seen allows arbitrarily large values of $\mathcal{A}$, may be misleading us.

When AdS-Kerr black holes are discussed, the specific structure of the metric tensor (see below) leads most authors to impose the condition
\begin{equation}\label{AA}
\mathcal{A}/L < 1.
\end{equation}
This is an unexpected restriction: it seems strange that the amount of angular momentum a black hole in AdS$_5$ can acquire should be strictly limited by the background asymptotic curvature, which does not appear to be relevant and which is in any case the dominant parameter only very far from the black hole.

In \cite{kn:96} we studied possible bounds on $\mathcal{A}/L$, focussing on a ``stringy'' effect, pointed out by Seiberg and Witten \cite{kn:seiberg,kn:maldacena,kn:KPR}, which arises as a possible pair-production instability (of branes) in asymptotically AdS spacetimes. The result was rather remarkable: avoidance of Seiberg-Witten instability does in fact translate in this case to an \emph{upper bound} on $\mathcal{A}/L$. That is, \emph{\emph{in string theory}}, a (singly rotating) AdS$_5$-Kerr black hole of given mass cannot rotate with arbitrarily large angular momentum; furthermore, the bound takes the simple form of bounding $\mathcal{A}/L$ by a universal constant. This effect arises because $\mathcal{A}$ and $L$ \emph{compete} to control the rate of growth of the areas and volumes enclosed by branes as one moves far away from the black hole; the instability is triggered when $\mathcal{A}$ ``outcompetes'' $L$. Thus string theory explains the fact that $\mathcal{A}$ must be bounded by (some multiple of) $L$ if the black hole is actually to exist within the theory.

The Seiberg-Witten upper bound on $\mathcal{A}/L$, however, is larger than unity (one finds $\mathcal{A}/L \leq \;\; \approx 2.83$); and so, in some cases, there is a gap between the lower bound imposed by classical Censorship and this upper bound. It appears that string theory does permit black holes with $\mathcal{A}/L > 1$, even if only in a narrow range.

We will argue that \emph{most} black holes corresponding to this ``gap'' are in fact unstable to the familiar \emph{superradiant instability} \cite{kn:super}. Nevertheless, even the condition that this instability should be absent does not quite eliminate all AdS$_5$-Kerr black holes with $\mathcal{A}/L > 1$: there remains a small set of parameter values, which we will identify precisely, such that the corresponding black holes are stable against both Seiberg-Witten and superradiant instabilities. We have been unable to exclude these on general grounds.

We begin with a brief review of the geometry and physical quantities describing a singly-rotating AdS$_5$-Kerr black hole.

\addtocounter{section}{1}
\section* {\large{\textsf{2. Singly Rotating AdS$_5$-Kerr Black Holes}}}
The AdS$_5$-Kerr metric \cite{kn:hawk,kn:cognola,kn:gibperry}, for a black hole rotating about a single axis, is given by
\begin{flalign}\label{A}
g\left(\m{AdSK}_5^{(a,0)}\right)\; = \; &- {\Delta_r \over \rho^2}\left[\,\m{d}t \; - \; {a \over \Xi}\,\m{sin}^2\theta \,\m{d}\phi\right]^2\;+\;{\rho^2 \over \Delta_r}\m{d}r^2\;+\;{\rho^2 \over \Delta_{\theta}}\m{d}\theta^2 \\ \notag \,\,\,\,&+\;{\m{sin}^2\theta \,\Delta_{\theta} \over \rho^2}\left[a\,\m{d}t \; - \;{r^2\,+\,a^2 \over \Xi}\,\m{d}\phi\right]^2 \;+\;r^2\cos^2\theta \,\m{d}\psi^2 ,
\end{flalign}
where
\begin{eqnarray}\label{B}
\rho^2& = & r^2\;+\;a^2\cos^2\theta, \nonumber\\
\Delta_r & = & \left(r^2+a^2\right)\left(1 + {r^2\over L^2}\right) - 2M,\nonumber\\
\Delta_{\theta}& = & 1 - {a^2\over L^2} \, \cos^2\theta, \nonumber\\
\Xi & = & 1 - {a^2\over L^2}.
\end{eqnarray}
Here $L$ is the AdS curvature length scale, $t$ and $r$ are as usual, ($\theta, \phi, \psi$) are Hopf coordinates on the three-sphere, and $a$ and $M$ are positive parameters with a \emph{purely geometric} meaning; however, they determine the physical mass $\mathcal{M}$, and the angular momentum per unit mass, $\mathcal{A}$, through the relations \cite{kn:gibperry}
\begin{equation}\label{C}
\mathcal{M}\;=\;{\pi M \left(2 + \Xi\right)\over 4\,\ell_{\textsf{B}}^3\,\Xi^2},
\end{equation}
\begin{equation}\label{D}
\mathcal{A}\;=\;{2 a \over 2 + \Xi}\;=\;{2 a \over 3 - \left(a^2/L^2\right)}.
\end{equation}
Here $\ell_{\textsf{B}}$ is the gravitational length scale in the bulk\footnote{In string theory, this gravitational length scale is related to the asymptotic AdS curvature scale $L$ by $\ell_{\textsf{B}}^3\; =\; {\pi\over 2}\times L^3 \times {1\over N_{\textsf{c}}^2},$ where $ N_{\textsf{c}}$ represents the number of colours describing the boundary field theory. This number is always assumed to be large, so we see that $\ell_{\textsf{B}}$ must be very much smaller than $L$.}. Notice that (\ref{C}) excludes $a = L$ (except possibly in some limiting sense, see \cite{kn:klem,kn:hen,kn:supe}). However, it does not in any way exclude $a > L$: see \cite{kn:96} for this case.

Notice that $\mathcal{M}$ depends on \emph{both} $M$ and $a$; notice too that, while there is a one-to-one correspondence between $a$ and $\mathcal{A}$, their ranges are very different: $a/L$ takes values in $[\,0,\,\sqrt{3}\,)$ (excluding unity, as above), while $\mathcal{A}/L$ ranges in $[\,0,\,\infty)$ (again excluding unity). It is useful to note that (\ref{D}) implies that $a/L < 1$ if and only if $\mathcal{A}/L < 1$.

Cosmic censorship for these black holes takes an unusual form \cite{kn:96}. First, we define a dimensionless physical mass of the black hole, $\mu$, by
\begin{equation}\label{E}
\mu \equiv {8\ell_{\textsf{B}}^3\mathcal{M}\over \pi L^2}.
\end{equation}
As usual, Cosmic Censorship is the condition that the equation $\Delta_r = 0$ (see (\ref{B}) above) should have physically acceptable solutions. Using the parameter $\Xi$ defined in (\ref{B}) above, one finds (using also (\ref{C}) and (\ref{D})) that the condition for Cosmic Censorship to hold in this case is a simple but \emph{quadratic} inequality,
\begin{equation}\label{XI}
\left(\mu + 1\right)\Xi^2\; + \;\Xi \; - 2 \;> \; 0.
\end{equation}
Because this is quadratic, there are two alternatives: Cosmic Censorship holds if and only if \emph{either}
\begin{equation}\label{F}
{\mathcal{A}\over L} \;<\; \Gamma_{\mu}^- \; \equiv \;2\,\sqrt{2}\,\sqrt{\mu + 1}\,{\sqrt{3 + 2\mu - \sqrt{9 + 8\mu}}\over 3 + 4\mu + \sqrt{9 + 8\mu}} \;<\;1,
\end{equation}
\emph{or}
\begin{equation}\label{G}
{\mathcal{A}\over L} \;>\; \Gamma_{\mu}^+ \;\equiv \; 2\,\sqrt{2}\,\sqrt{\mu + 1}\,{\sqrt{3 + 2\mu + \sqrt{9 + 8\mu}}\over 3 + 4\mu - \sqrt{9 + 8\mu}}\;>\;1.
\end{equation}
In this case, then, classical Censorship does not forbid arbitrarily large angular momenta per unit mass, as it does in the asymptotically flat case: it only excludes a \emph{band} of values of $\mathcal{A}/L$ around unity (for a given fixed physical mass). The band becomes progressively narrower for larger values of the mass.

Our interest here is in what happens when these objects are embedded in a more complete theory, which we take to be string theory. Let us now consider that.

\addtocounter{section}{1}
\section* {\large{\textsf{3. The Seiberg-Witten Criterion}}}
Seiberg and Witten \cite{kn:seiberg} (see \cite{kn:maldacena,kn:KPR}) showed that the study of BPS branes propagating in asymptotically AdS spacetimes leads to a condition which must be satisfied if the system is to be stable. Subsequent work \cite{kn:ferrari1,kn:ferrari2,kn:ferrari3,kn:ferrari4} has shown that this condition is actually a reflection of very deep internal consistency conditions in string theory; systems which fail to satisfy it \emph{are not solutions in the full theory}.

Following \cite{kn:seiberg}, in \cite{kn:96} we considered a BPS brane located at radial coordinate value $r$ in the singly rotating AdS$_5$-Kerr spacetime, and computed the action. For large $r$, one finds that this action takes the form
\begin{equation}\label{H}
\textsf{S}\;\propto\;\left(1 - {a^2\over 2L^2}\right)r^2 + C + {D\over r^2} + \cdots,
\end{equation}
where $C$ and $D$ are constants determined by the black hole parameters; the ellipsis denotes terms which are negligible at large $r$; the constant of proportionality is positive. Clearly, the action will be unbounded below, signalling an instability, unless the condition
\begin{equation}\label{I}
a/L \; \leq \; \sqrt{2}
\end{equation}
is satisfied. In terms of the more physical parameter $\mathcal{A}$, we must have
\begin{equation}\label{J}
\mathcal{A}/L \; \leq \; 2\sqrt{2}
\end{equation}
if the black hole is to be stable in string theory. We will refer to either (\ref{I}) or (\ref{J}) as the \emph{Seiberg-Witten criterion}.

Thus indeed \emph{string theory forbids arbitrarily large values of} $\mathcal{A}/L$. As we explained earlier, this is precisely in line with general holographic expectations.

Note that, from (\ref{F}), $\Gamma_{\mu}^+ \leq 2\sqrt{2}$ if and only if $\mu \geq 2$. That is, for all values of $\mu$ strictly greater than 2, there is a \emph{gap} between the lower bound on $\mathcal{A}/L$ imposed by Cosmic Censorship, and the upper bound on it imposed by the Seiberg-Witten condition.

More precisely, we have, in the case $a/L > 1$, $\mu > 2$,
\begin{equation}\label{K}
1\;<\;\Gamma_{\mu}^+\;<\;{\mathcal{A}\over L} \;\leq\; 2\sqrt{2}.
\end{equation}

In terms of the $(M/L^2, a/L)$ parameter space for these black holes, we have so far the following constraints.

First, for each value of $a/L$, Cosmic Censorship imposes a lower bound on $M/L^2$: this is shown as the leftmost, nearly vertical curve in Figure 2 below. Only those points to the right of that line are permitted.

Second, we have just seen that the Seiberg-Witten criterion rules out values of $a/L$ larger than $\sqrt{2}$. This is shown as the top, horizontal line in Figure 2 below. Only those points below that line are permitted.

In summary: string theory disallows an infinite range of values for $\mathcal{A}/L$, otherwise permitted by classical Censorship. However, it does still apparently allow \emph{some} values of $\mathcal{A}/L > 1$. It is natural to ask whether the gap permitted by (\ref{K}) actually exists in a physical sense, and, if so, what parameter values are allowed. In the next Section, we consider this.

\addtocounter{section}{1}
\section* {\large{\textsf{4. Superradiance When $\mathcal{A}/L > 1.$}}}
As is well known, rapidly rotating AdS black holes are at risk of being destabilized by the emission of \emph{superradiant modes} \cite{kn:super}: these cause the black hole to lose both mass and angular momentum. Therefore, if the condition for stability against superradiance is not satisfied, the black hole must evolve until the parameters are such that it \emph{is} satisfied.

It is natural to suspect that very rapidly rotating AdS$_5$-Kerr black holes, with $\mathcal{A}/L > 1$, might all be unstable in this manner. If that were the case, then the black hole would evolve until $\mathcal{A}/L < 1$ and superradiance ceases. (Of course, the corresponding time-dependent metric is quite different from the AdS$_5$-Kerr metric, and it need not be subject to the constraints necessarily satisfied by the latter; in particular, the prohibition on $\mathcal{A}/L = 1$ need not apply. That is, superradiance cannot be avoided by invoking that prohibition.)

In our case, the relevant quantity is given by
\begin{equation}\label{L}
\varsigma \;=\; {{a\over L}\,\left(1 + {r_H^2\over L^2}\right)\over {r_H^2\over L^2} + {a^2\over L^2}}.
\end{equation}
It is known \cite{kn:reall} that a sufficient condition for stability against superradiance is that $\varsigma < 1$, and there is good evidence that this condition is also necessary; here we will assume that to be the case.

Clearly $\varsigma$ can be regarded as a function of $a/L$ and $r_H/L$. However, the latter is given in terms of $a/L$ and $M/L^2$ through solving the equation
\begin{equation}\label{M}
\left(r_H^2+a^2\right)\left(1 + {r_H^2\over L^2}\right) - 2M\;=\;0,
\end{equation}
derived from the second relation in the set (\ref{B}). Thus we think of $\varsigma$ as a function of $a/L$ and $M/L^2$: as such, its graph is shown as Figure 1.

\begin{figure}[!h]
\centering
\includegraphics[width=1.1\textwidth]{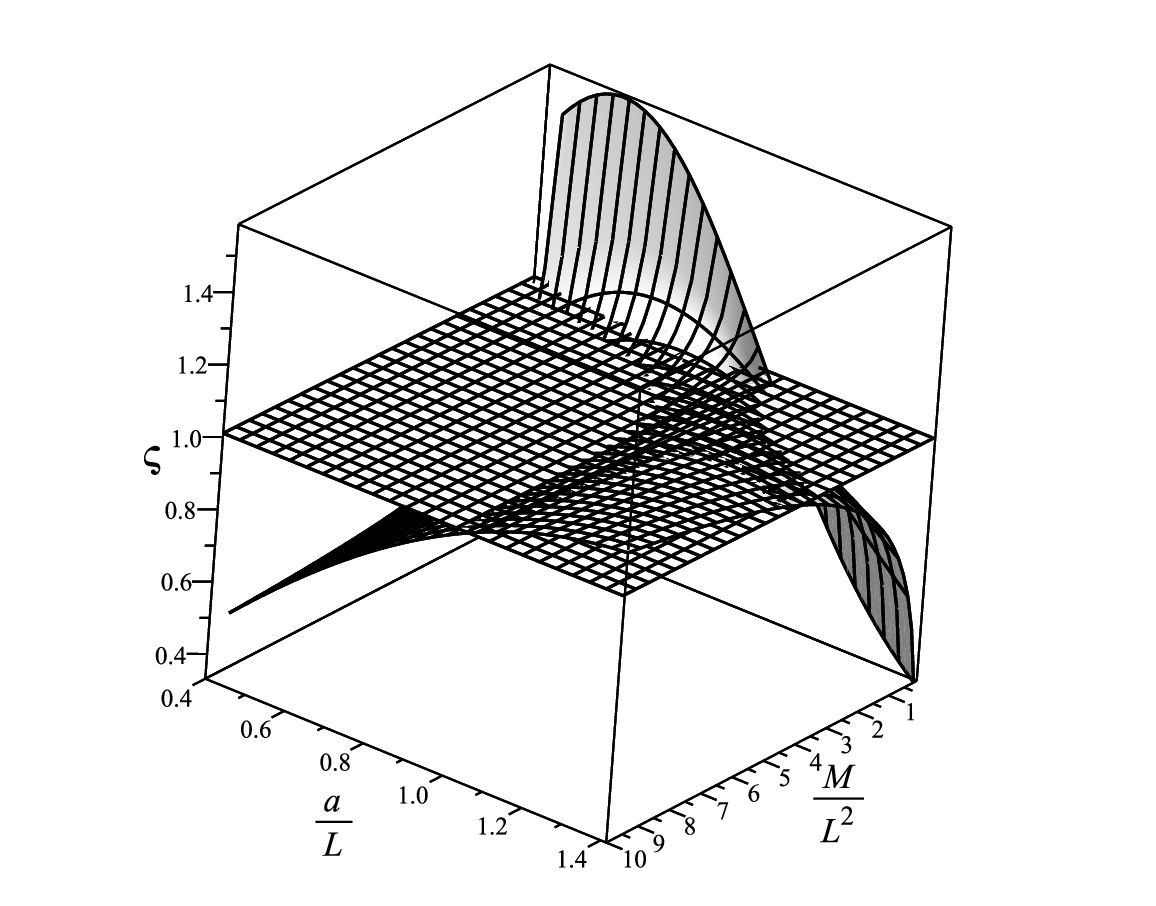}
\caption{Graph of $\varsigma(M/L^2, a/L)$. (The plane $\varsigma = 1$ is also shown to ease interpretation.) Notice that when $a/L < 1,$ this function is below unity only for relatively large values of $M/L^2$; whereas, when $a/L > 1,$ the function is below unity only for relatively small values of $M/L^2$.}
\end{figure}

One sees from this graph that, when $a/L < 1$, superradiance can be avoided only by taking $M/L^2$ to be relatively \emph{large}. On the other hand, it is clear that superradiance can be avoided when $a/L > 1$ only by taking $M/L^2$ to be relatively \emph{small}, that is, bounded above. Thus we see that requiring the absence of superradiance when $a/L > 1$ has the dramatic effect of cutting off an infinite region in the parameter space of these black holes: only those points in Figure 2 which lie to the left of the rightmost, nearly vertical curve, are acceptable.

The precise form of that curve is given by solving the equation $\varsigma(M/L^2, a/L) = 1$; this gives us
\begin{equation}\label{N}
a/L\;=\;{1\over 3}\left(1+27{M\over L^2}+3\sqrt{81{M^2\over L^4}+6{M\over L^2}}\right)^{1/3}+{1\over 3\left(1+27{M\over L^2}+3\sqrt{81{M^2\over L^4}+6{M\over L^2}}\right)^{1/3}}\,-\,{2\over 3},
\end{equation}
and this is the curve shown in Figure 2.

In summary: \emph{It is possible for an AdS$_5$-Kerr black hole with $a/L > 1$ to satisfy Cosmic Censorship and be stable against both stringy and superradiant instablities}. But this is only possible for black holes with parameters inside the small parameter domain portrayed in Figure 2.
\begin{figure}[!h]
\centering
\includegraphics[width=0.7\textwidth]{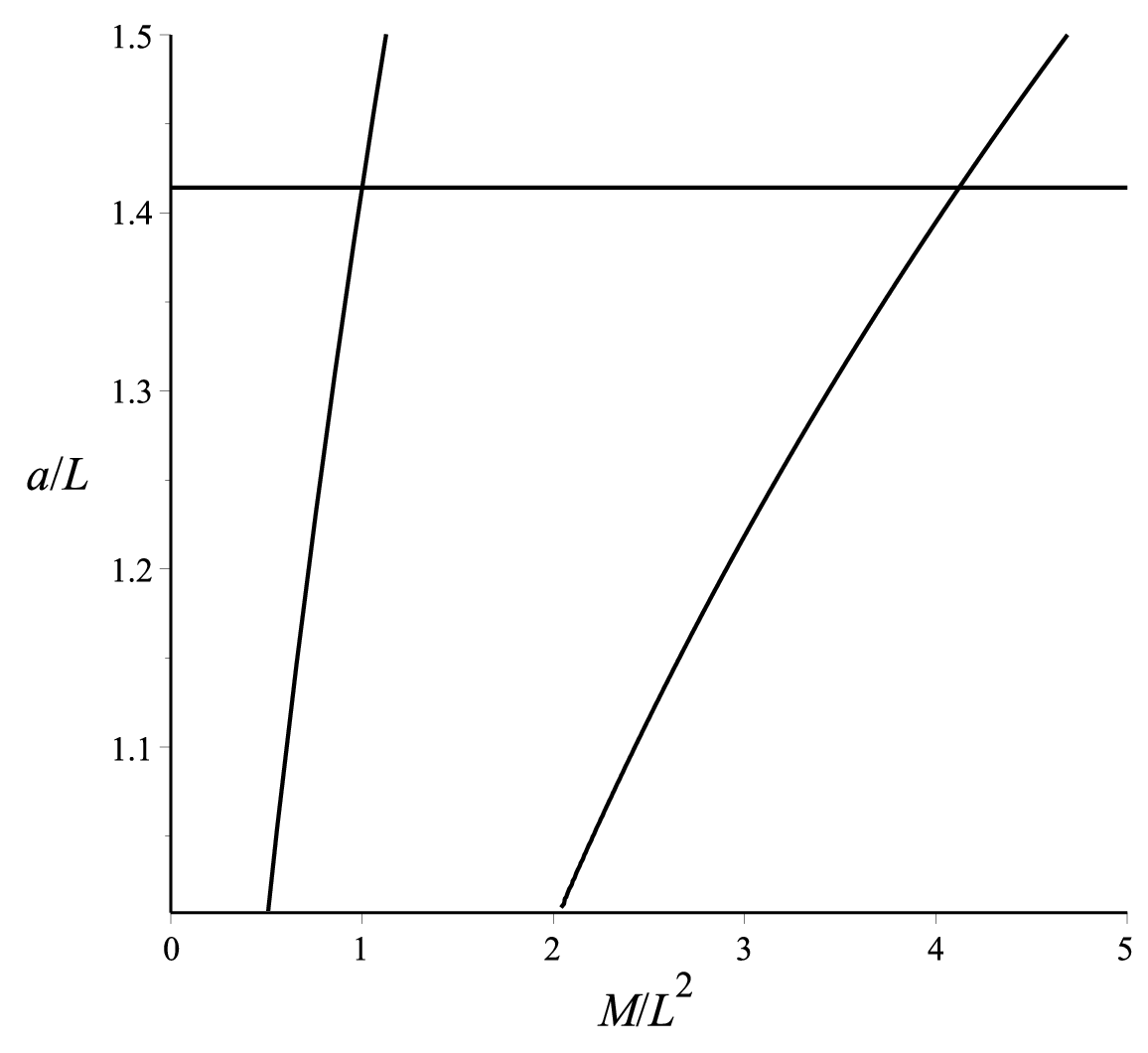}
\caption{Stable AdS$_5$-Kerr black holes with $a/L > 1$ have $(M/L^2, a/L)$ values bounded by lines representing (in clockwise order from left) Censorship, the Seiberg-Witten condition, and stability against superradiance. The surviving black holes correspond to the domain with vertices, in clockwise order, $(M/L^2 , a/L) = (0.5, 1), (1, \sqrt{2}), ((3\sqrt{2} + 4)/2 \approx 4.121, \sqrt{2}), (2, 1)$.}
\end{figure}

We see, in particular, that the value of $M/L^2$ for any point in this domain always exceeds\footnote{In \cite{kn:klem,kn:hen,kn:supe} an attempt is made to construct black holes with finite mass and with $a \rightarrow L$. One sees from equation (\ref{C}) that this is only possible here by taking $M$ arbitrarily small. From our discussion here, we see that this means that the $a \rightarrow L$ limit cannot be taken from above.} $0.5$, while the largest possible value is $((3\sqrt{2} + 4)/2 \approx 4.121$. \emph{No AdS$_5$-Kerr black hole with $\mathcal{A}/L > 1$ and $M/L^2$ outside these narrow ranges can satisfy Censorship and be stable}. This is the analogue of the inequalities given in (\ref{K}).
	
\addtocounter{section}{1}
\section* {\large{\textsf{5. Conclusion }}}

Classical censorship allows an AdS$_5$-Kerr black hole to have an arbitrarily large angular momentum for a given mass. String theory, however, strongly curtails this freedom, imposing in fact a universal upper bound on $\mathcal{A}/L$.

Our study of superradiance for these objects leads to a surprising conclusion: AdS$_5$-Kerr black holes with $\mathcal{A}/L$ values beyond unity, indeed all the way up to the stringy (Seiberg-Witten) bound $2\sqrt{2}$ \emph{can} exist as stable objects. However, this is only possible for a small domain in the AdS$_5$-Kerr parameter space. These very special black holes evidently merit closer attention; one is naturally led to ask: perhaps they are unstable in some other way? We hope to return to this question elsewhere.

\addtocounter{section}{1}
\section*{\large{\textsf{Acknowledgements}}}
The author is grateful to Prof. Ong Yen Chin for drawing his attention to useful references, and to Dr. Soon Wanmei for helpful discussions.

\end{document}